\documentclass[fleqn,usenatbib]{mnras}

\usepackage{newtxtext,newtxmath}

\usepackage[T1]{fontenc}

\DeclareRobustCommand{\VAN}[3]{#2}
\let\VANthebibliography\thebibliography
\def\thebibliography{\DeclareRobustCommand{\VAN}[3]{##3}\VANthebibliography}

\usepackage{graphicx}	
\usepackage{amsmath}	
\usepackage{newtxmath}
\usepackage{soul}
\usepackage{float}
\usepackage{dsfont}
\usepackage{subfigure}
\usepackage{comment}
\usepackage{color}
\usepackage[normalem]{ulem}
\usepackage[dvipsnames]{xcolor} 

\title[GW200105 and GW200115]{Modelling the formation of the first two neutron star-black hole mergers, GW200105 and GW200115: metallicity, chirp masses and merger remnant spins}

\author[D. Chattopadhyay et al.]{Debatri Chattopadhyay,$^{1,2,3}$\thanks{E-mail: ChattopadhyayD@cardiff.ac.uk}
Simon Stevenson,$^{1,2}$
Floor Broekgaarden,$^{4}$
Fabio Antonini,$^{3}$
\newauthor
Krzysztof Belczynski$^{5}$\\
\\
$^{1}${Centre for Astrophysics and Supercomputing, Swinburne University of Technology, John St., Hawthorn, Victoria- 3122, Australia}\\
$^{2}${The ARC Centre of Excellence for Gravitational Wave Discovery,  OzGrav}\\
$^{3}${Gravity Exploration Institute, School of Physics and Astronomy, Cardiff University, Cardiff, CF24 3AA, UK}\\
$^{4}${Center for Astrophysics \textbar{} Harvard $\&$ Smithsonian,
60 Garden Street, Cambridge, MA 02138, USA}\\
$^{5}${Nicolaus Copernicus Astronomical Center, Polish Academy of Sciences, ul. Bartycka 18, 00-716 Warsaw, Poland}
}

\date{Accepted XXX. Received YYY; in original form ZZZ}

\pubyear{2015}

\begin{document}
\label{firstpage}
\pagerange{\pageref{firstpage}--\pageref{lastpage}}
\maketitle

\begin{abstract}
The two neutron star-black hole mergers (GW200105 and GW200115) observed in gravitational waves by advanced LIGO and Virgo, mark the first ever discovery of such binaries in nature. We study these two neutron star-black hole systems through isolated binary evolution, using a grid of population synthesis models. 
Using both mass and spin observations (chirp mass, effective spin and remnant spin) of the binaries, we probe their different possible formation channels in different metallicity environments. 
Our models only support LIGO data when assuming the black hole is non spinning.
Our results show a strong preference that GW200105 and GW200115 formed from stars with sub-solar metallicities $Z\lesssim 0.005$.
Only two metal-rich ($Z=0.02$) models are in agreement with GW200115. 
We also find that chirp mass and remnant spins jointly aid in constraining the models, whilst the effective spin parameter does not add any further information. We also present the observable (i.e. post selection effects) median values of spin and mass distribution from all our models, which maybe used as a reference for future mergers. Further, we show that the remnant spin parameter distribution exhibits  distinguishable features in different neutron star-black hole sub-populations. We find that non-spinning, first born black holes dominate significantly the merging neutron star-black hole population, ensuring electromagnetic counterparts to such mergers a rare affair.
\end{abstract}




\section{Introduction}
\label{sec:intro}

The discovery of the two merger events GW200105 and GW200115 by the Advanced LIGO \citep{TheLIGOScientificDetector:2014jea} and Virgo \citep{TheVirgoDetector:2014hva} gravitational-wave detectors marks the first observations of a black hole-neutron star binary \citep{LIGOScientific:2021qlt}. 
While there have been predictions of black hole-neutron star (NS+BH) systems before---accounting for their non-detection as well as future possibilities of detection \citep{Siggurdson:2003, Clausen:2014, Chattopadhyay:2020lff, Broekgaarden:2020NSBH}---these two events provide the perfect opportunity to re-assess our understanding of the formation and evolution of massive binaries towards NS+BH systems.

The properties of GW200105 and GW200115 are consistent with astrophysical expectations for NS+BH binaries.
Under a conservative prior allowing for large spins ($<0.99$) for the putative neutron star, GW200105 is formed through the merger of a $8.9^{+1.2}_{-1.5}$\,M$_\odot$\footnote{90\% credible interval, from \citet{LIGOScientific:2021qlt}}  black hole (BH) with a $1.9^{+0.3}_{-0.2}$\,M$_\odot$ neutron star (NS), whilst GW200115 was through the merger of $5.7^{+1.8}_{-2.1}$\,M$_\odot$ BH with a $1.5^{+0.7}_{-0.3}$\,M$_\odot$ NS. 
Under a more constraining spin prior with low neutron star spins ($<0.05$), informed by observations of Galactic NSs,  GW200105 and GW200115 are inferred to have formed through the coalescences of $8.9^{+1.2}_{-1.5}$\,M$_\odot$ BH with a $1.9^{+0.2}_{-0.2}$\,M$_\odot$ NS and $5.9^{+1.4}_{-2.1}$\,M$_\odot$ BH with a $1.4^{+0.6}_{-0.2}$\,M$_\odot$ NS respectively. 
The effective spin ($\chi_{\mathrm{eff}}$) for GW200105 is found to be $-0.01^{+0.11}_{-0.15}$ ($-0.01^{+0.08}_{-0.12}$), while for GW200115 it is $-0.19^{+0.23}_{-0.35}$ ($-0.14^{+0.17}_{-0.34}$) for high (low) secondary spin priors respectively. 
The mass and spin ($\chi_{\mathrm{rem}}$) of the post-merger remnant BH for GW200115 are obtained to be  $10.4^{+2.7}_{-2.0}$\,M$_\odot$ and $0.43^{+0.04}_{-0.03}$ respectively. 
For GW200105, the remnant BH mass is $7.8^{+1.4}_{-1.6}$\,M$_\odot$ and $\chi_{\mathrm{rem}} = 0.38^{+0.04}_{-0.02}$. Rather than individual masses, mass ratio or individual spins, the chirp mass, effective and remnant spins are better constrained observable quantities.

The two NS+BH gravitational wave (GW) observations inspire us to ask some fundamental questions about their formation scenario. 
\citet{Broekgaarden:2021hlu} have shown that the masses and merger rates inferred from GW200105 and GW200115 are consistent with formation through isolated binary evolution. 
In particular, they explored a large set of binary evolution models, and found that low kick velocities or relatively high common envelope efficiencies are preferred to simultaneously match the properties of double NS and double BH systems. In this study, we perform population synthesis simulations to investigate the birth order of the two types of compact objects - NS and BH - in the NS+BH systems, their masses, spins and expected merger rates. 

Particularly, we address the following primary questions in this study ---\\
a) can the observational mass and spin signatures aid in distinguishing which sub-population -- if the BH is born first (BHNS) or the NS is born first (NSBH)\footnote{NSBH systems are exciting especially as multi-messenger sources --- they can harbour recycled pulsars \citep{Chattopadhyay:2020lff} as the NSs is born first and NSBHs with high pre-merger BH spins have the possibility to produce exciting electromagnetic signatures at their mergers \citep{Hu:2022ubh}} -- GW200105 and GW200115 were a part of? \\
b) what are the overall predicted rates of BHNS vs NSBH formation across different metallicities? \\
c) Did the pre-merger BHs for both the observations have a considerable spin ($\chi_\mathrm{BH} > 0.1$)?\\
d) Does the observed remnant BH spins of GW200105 and GW200115 add information that can help constrain population synthesis models?\\
e) can the formation metallicity of the two NS+BH merger signals be inferred  from the qualitative analysis of the observations?\\

In this paper, we use similar models to \citet{Broekgaarden:2021hlu}, but also compare measurements of the compact objects' spins to those predicted from our model. We assume first born BH to be non-spinning in all models and explore submodels in which the second born BH can either be spinning or non-spinning. 
For all but one of our models we assume that the spins of the component black holes are aligned with the binary orbital angular momentum prior to merger. We additionally test if spin misalignment due to supernova kicks heavily impacts our results.
Using this additional information, we aim to constrain whether the neutron star or black hole formed first in GW200105 and GW200115, and if we can place constraints on the metallicity of the environment in which they were born.
We find in our study that though we cannot draw any additional constraints on binary evolution models, we do get an idea of the possible formation metallicities of these binaries through the analysis of the remnant spin in addition to chirp mass.

This paper is structured as follows: in section~\ref{sec:methods} we outline the population synthesis methods we use to evolve isolated binaries; in section~\ref{sec:results} we discuss the results of our models by comparing them to the chirp mass and remnant spin observations of GW200105 and GW200115;  in section~\ref{sec:conclusions} we conclude and discuss our findings (including the answers to the five key questions outlined above) and their implications.

\section{Methods}
\label{sec:methods}

In this paper, we study the possible formation of the two NS+BH observations through isolated binary evolution, using the population synthesis code COMPAS \citep{Stevenson:2017tfq,Vigna-Gomez:2018dza,COMPAS:2021methodsPaper}. COMPAS is based on updated stellar and binary evolution prescriptions from \citet{SSE_Hurley:2000pk} and \citep{BSE:2002} respectively. 
There are many uncertainties in massive binary evolution \citep[see for example][]{Klencki:2018, Giacobbo:2018, Belczynski:2021zaz,  Broekgaarden:2020NSBH, Chattopadhyay:2020lff}. 
To investigate what we can learn from GW200105 and GW200115, we have produced a set of four main models (``Pessimistic", ``Optimistic", ``Kick\_100", ``Alpha\_2") with different stellar/binary evolution assumptions. 
``Pessimistic" is our base model, with one key assumption about binary or stellar evolution varied in each of the following three models. 
In our choice of parameters for the models, we are guided by Fiducial/Default models of previous population synthesis studies \citep{Stevenson:2017tfq,Vigna-Gomez:2018dza,Neijssel:2019,Chattopadhyay:2019xye,Chattopadhyay:2020lff,Broekgaarden:2020NSBH}.
Based on the recent results from \citet{Broekgaarden:2021hlu}, we place particular emphasis on the importance of neutron star natal kicks, and the efficiency of the common envelope phase of binary evolution.
\citet{Broekgaarden:2021hlu} showed that natal kick velocities less than 100\,km/sec or high common envelope efficiency parameter ($\alpha_\mathrm{CE} \gtrsim 2$) provide good match to the merger rates of double BH, double NS and NS+BH inferred from gravitational waves observations \citep{LIGOScientific:2021psn}.  
Many processes in binary evolution are sensitive to metallicity. 
To examine the impact of metallicity (and to attempt to constrain the formation metallicities of GW200105 and GW200115), we evolve each of the four models at four different metaillicites ($Z$). 
From lower to higher metallicity, we have models M-005 ($Z=0.0005$), M-01 ($Z=0.001$), M-05 ($Z=0.005$), M-2 ($Z=0.02$).

The zero age main sequence (ZAMS) mass of the primary (initially more massive) star is drawn from an initial mass function \citep{Kroupa:2000iv}, while the mass of the secondary is determined by drawing the mass ratio of the binary following an uniform distribution \citep{Sana:2012}. All binaries at assumed to have a circular orbit at ZAMS \citep{BSE:2002}
The initial semi-major axes of the binaries 
follow the \cite{Sana:2012} distribution, with the minimum and maximum of the distribution\footnote{The orbital period $p$ in days is given by $f(\log{p}) \propto (\log{p})^\pi$, where $\pi=-0.55$. The semi-major axis is calculated from $p$ by using Kepler's third law for circular orbits.} being 0.1 AU to 1000 AU respectively. 
Variations in the distributions chosen for the initial conditions do not typically have a large impact on results for NS+BH binaries \citep[e.g.,][]{deMink:2015yea}

In the dominant channel of NS+BH formation, during Roche lobe overflow (RLOF), the mass transfer efficiency $\beta_\mathrm{MT}$, which is the ratio of mass accreted by the accretor and the mass lost by the donor is given by,
\begin{equation}
    \beta_\mathrm{MT} = \mathrm{min}(1, 10 \frac{\tau_\mathrm{acc}}{\tau_\mathrm{don}}) \, ,
    \label{eq:beta_thermal}
\end{equation}
\citep{BSE:2002,  Schneider:2015ApJ...805...20S} where the thermal (Kelvin-Helmholtz) timescales for the accretor and donor are denoted by $\tau_\mathrm{acc}$ and $\tau_\mathrm{don}$ respectively \citep[see][equation 2 for futher details]{Chattopadhyay:2020lff}. 
Mass accretion onto a compact object is assumed to be Eddington limited \citep{Stevenson:2017tfq,Chattopadhyay:2019xye}. 

The common envelope (CE) phase that may arise from unstable mass transfer is assumed to have $\alpha_\mathrm{CE} = 1$, where $\alpha_\mathrm{CE}$ denotes the fraction of the orbital energy available to unbind the envelope. 
The binding energy of the envelope is determined using the fits to detailed models from \citet{XuLi:2010}. 
For model ``Alpha\_2", we varied this efficiency parameter to $\alpha_\mathrm{CE}=2$.

We assume the `pessimistic' 
scenario for common envelope evolution, where a Hertzsprung gap (HG) donor is not allowed to survive CE and instead merges \citep{Belczynski:2007ApJ}. 
In the ``Optimistic" model we alter this assumption, allowing for the possibility for HG donors to survive CE evolution and successfully eject the envelope. 
For all models, a NS companion is allowed to accrete during CE following \citet{MacLeod:2014yda}, as described in details in \citet{Chattopadhyay:2019xye}. 
Since Pessimistic, Kick\_100 and Alpha\_2 are all under the `pessimistic' CE framework, in this paper we sometimes refer to all three as `pessimistic'.

For a more comprehensive description of COMPAS prescriptions of mass transfer during RLOF and common envelope, see the COMPAS methods paper \citep{COMPAS:2021methodsPaper}, and specifically for the case of NS+BH we refer to \citet{Chattopadhyay:2020lff,Broekgaarden:2020NSBH,Broekgaarden:2021hlu}. 

In COMPAS, we assume a Maxwellian supernova natal kick distribution for NSs, with the one dimensional root mean square $\sigma_\mathrm{CCSN}$=265\,km/sec for core collapse supernovae \citep{Hobbs:2005}, $\sigma_\mathrm{ECSN}$=$\sigma_\mathrm{USSN}$=30\,km/sec for electron capture and ultra-stripped supernovae \citep{Podsiadlowski:2004, Gessner:2018ekd, Muller:2018utr}. 
The assumption for kick velocity is varied in model ``Kick\_100", where  $\sigma_\mathrm{CCSN}$ is assumed to be 100\,km/sec.
The BH natal kicks are also drawn from identical distributions as NS, but are scaled down by the fallback mass fraction \citep{Fryer:2012}. 

NSs are also treated as pulsars in COMPAS, with spins and surface magnetic fields computed in isolated (no mass transfer) and mass transfer (during RLOF and CE) cases, as detailed in \citet{Chattopadhyay:2019xye}. 
Natal spins of the NSs are drawn from an uniform distribution (between 10--100 milliseconds) and they are expected to spin down rapidly $\sim\mathcal{O}(10-100$\,Myrs). 
Angular momentum transfer during mass accretion can spin pulsars up as well. 
\citet{Chattopadhyay:2020lff} estimated that the mass weighted spin $\chi_\mathrm{NS}\lesssim$0.3 for solar metallicity. 

Due to efficient angular momentum transfer from the core to the envelope of BH progenitors, first born BHs are assumed to be non-spinning in all our models \citep{Fuller:2019sxi}. 
The progenitor of the second born BH, however, may get spun up through tidal interaction with its compact object companion (in the case of this paper, a NS), depending on the orbital configuration of the binary \citep{Qin:2018nuz, Bavera:2019, Chattopadhyay:2020lff, Hu:2022ubh}. 
We assume two models - i) BH-Q: where the second born BH is allowed to be spun up  as guided through prescriptions for different metallicities from \citet{Qin:2018nuz} and ii) BH-Z: where BHs produced through second supernovae are also non-spinning. 
While \citet{Chattopadhyay:2020lff} focused on the pre-merger BH spin analysis of only solar metallicity $Z=0.0142$ \citep{Asplund:2009} models (in that paper, we were more interested in the Galactic population), in this paper we explore a grid of four metallicities ($Z=0.02,0.005,0.001,0.0005$).
For BH-Q models, the pre-merger, second-born BH spin $\chi_\mathrm{BH}$ is given as, 
\begin{equation}
    \chi_{\mathrm{BH}} = \left\{\begin{array}{ll}
                    0, & \text{for } \log_{10}{P_\mathrm{orb}} >  x_1 \\
                    1, & \text{for } \log_{10}{P_\mathrm{orb}} < x_2 \\
                    m \log_{10} P_\mathrm{orb} + c, & \text{for } x_2 \le \log_{10}{P_\mathrm{orb}} \le x_1
                    \end{array}\right\}
    \label{eq:aBHfitQin}
\end{equation}
where,
\begin{equation}
    x_1; x_2; m; c = \left\{\begin{array}{ll}
                    0.5; -0.5; -0.87; 0.57;& \text{for } Z = 0.02 \\
                    0.5; -0.5; -0.70; 0.54;& \text{for } Z = 0.005\\
                    0.5; -0.5; -0.70; 0.54;& \text{for } Z = 0.001 \\
                    0.3; -0.5; -1.02; 0.63;& \text{for } Z = 0.0005 \\
                    \end{array}\right\}
    \label{eq:params}
\end{equation}
for different metallicities (using fits from \citet{Qin:2018nuz}). 

We compute the effective spin parameter $\chi_\mathrm{eff}$ assuming that the spin of the NS ($\chi_\mathrm{NS}$) and the spin of the BH ($\chi_\mathrm{BH}$) are both completely aligned with the orbital angular momentum. 

This might be the case if  binary evolution processes such as tides and mass 
accretion (including during CE 
will have aligned the spins of the stars in the binary with the orbit, prior to the formation of the compact objects.
Small kicks associated with the formation of the compact objects that do not disrupt the binary would not be expected to reintroduce significant misalignments \citep{Stevenson:2017tfq}\footnote{There can be infrequent cases of mass transfer anti-aligning the spin of the system studied for the case of double BHs by \citet{Stegmann:2020kgb}. However, such amount of mass loss from the donor is expected to be statistically insignificant for NS+BHs than double BHs.}. 
The effective spin
\begin{equation}
    \label{eq:chi_effec}
    \chi_\mathrm{eff} = \frac{m_\mathrm{BH}\chi_\mathrm{BH} \cos\theta_\mathrm{BH}+m_\mathrm{NS}\chi_\mathrm{NS} \cos\theta_\mathrm{NS}}{m_\mathrm{BH}+m_\mathrm{NS}} ,
\end{equation}
from \citet{Cutler:1993}, where the misalignment angles $\theta_\mathrm{BH}$ and $\theta_\mathrm{NS}$ are both zero for all four COMPAS models.

To check if supernovae kicks can create significant misalignment in the binaries to account for a negative effective spin, we separately analyse a Pessimistic model simulated with StarTrack \citep{Belczynski:2005mr} -- called `PessimisticST' -- as a special sanity check run - that also tracks the binary spin misalignment. PessimisticST allow inefficient tidal interaction that may not completely align the systems with respect to orbital angular momentum at the time of merger (see \citealp{Olejak:2021} for details). 
Since StarTrack does not yet compute the spin evolution of NSs, the pre-merger $\chi_\mathrm{NS}$ was taken as 0.01 for all systems (guided by Fig.21 of \citep{Chattopadhyay:2020lff}).
Although accounting for misalignment due to supernovae kicks, this model does not change our final conclusions. 

The remnant black holes formed in NS+BH binary mergers are expected to be spinning.
Qualitatively, the post-merger remnant spin depends on the mass ratio of the binary prior to merger, 
the spin of the BH and the NS quadrupole tidal polarizability $\Lambda$ \citep{Zappa:2019ntl}.
The latter is a function of NS compactness which depends on the equation of state of the NS \citep{Damour:2009wj, Hinderer:2010}, and hence its mass-radius relation. 
The primary contribution to the remnant BH spin is from the orbital angular momentum of the binary, which is determined by the mass ratio of the binary. 

It is expected the NS would usually be either severely deformed or destructed due to tidal disruption in a few orbits before merger, or can also merger head-on with the BH if the innermost stable orbit is located within the radius of the BH \citep{Lattimer:1976, Hinderer:2010,Foucart:2020review}. 
Current numerical models assume a non-spinning pre-merger NS and this assumption, though not correct, is still expected to produce reasonable results \citep{Kochanek:1992ApJ, Zappa:2019ntl}. 

If the mass ratio of the NS+BH binary is $q_\mathrm{s}=M_\mathrm{BH}/M_\mathrm{NS}$ and its symmetric mass ratio is $\gamma = q_s/(1+q_s)^2$, using fits from \citet{Zappa:2019ntl} we obtain the remnant BH spin 
\begin{equation}
  \chi_{\mathrm{rem}}= a_4\times\gamma^4 + a_3\times\gamma^3 + a_2\times\gamma^2 + a_1\times\gamma^1 + a_0 ,
  \label{equ:chi_remnant}
\end{equation}
where
\begin{equation}
\begin{array}{l}
  a_4 = -2310.4\times\chi_{\mathrm{BH}}^2 + 2088.4\times\chi_{\mathrm{BH}} - 400 , \\
  \\
  a_3 = 1582.08\times\chi_{\mathrm{BH}}^2 - 1417.68\times\chi_{\mathrm{BH}} + 253.33 , \\
  \\
  a_2 = -367.46\times\chi_{\mathrm{BH}}^2 + 325.37\times\chi_{\mathrm{BH}} - 54.99 , \\
  \\
  a_1 = 34.43\times\chi_{\mathrm{BH}}^2 - 32.83\times\chi_{\mathrm{BH}} + 7.56 , \\
  \\
  a_0 = -1.17\times\chi_{\mathrm{BH}}^2 + 1.95\times\chi_{\mathrm{BH}} - 0.1 ,
  \label{equ:chi_remnant_coefficients}
  \end{array}
\end{equation}
as was determined in \citet{Chattopadhyay:2020lff}, under the assumption of a constant $\Lambda$ for all NSs, appropriate given that we assume all neutron stars have a radius of 12\,km.

We noticed in our analysis that although there were instances of models deemed as accepted/potential match by $\chi_\mathrm{eff}$ (solely or jointly with $M_\mathrm{chirp}$) were rejected due to mismatch with $\chi_\mathrm{rem}$, there were no instances of models accepted/potential match by $\chi_\mathrm{rem}$ that were rejected due to inconsistency with $\chi_\mathrm{eff}$. 

We also mention here that albeit $M_\mathrm{chirp}$ and $\chi_\mathrm{rem}$ are both partial functions of the binary mass ratio, and mass ratio being a more intrinsic binary property, it has high uncertainties observationally. Moreover, for spinning pre-merger BH model (NSBHs, where it may be $\chi_\mathrm{BH}>0$) $\chi_\mathrm{rem}$ no longer solely remains dependant on mass ratio but also on pre-supernovae orbital properties. Similarly, individual masses of the BH and the NS are less  constrained than $M_\mathrm{chirp}$. Hence we take the observationally more prominent signatures of the binary ($M_\mathrm{chirp}$ and $\chi_\mathrm{rem}$) in our comparisons than the intrinsic binary evolution properties (like individual masses or mass ratio).

Stellar metallicity is known to play an important role in the masses of the compact objects, specifically the BHs, through determining wind mass loss for massive ($\gtrsim 30$\,M$_\odot$) stars \citep{Vink:2001, Belczynski:2010, Mapelli:2010}. As stellar winds also increase the orbital separation in binaries, the mass transfer history of the binary is also affected by their metallicity. 
In general NS+BHs formed at lower metallicity not only tend to be more massive, but their orbital properties, and hence merger timescales and spin properties also depend on $Z$.

We evolve 10$^6$ isolated binaries for each model (Pessimistic, Optimistic, Alpha\_2, Kick\_100) at each metallicity $Z=0.0005,0.001,0.005,0.02$, resulting in 16 individual models. 
We further sub-select systems that merge in 13\,Gyrs (a Hubble Time, HT) for our analysis as a cut-off, since any system that takes longer to merge will not be a part of the LIGO observations.

We did not account for the redshift evolution of the binaries for the following reasons. The metallicity evolution with redshift is largely uncertain \citep{Furlong:2015, Neijssel:2019}, and even the local universe has low metallicity regions \citep{Tramper:2014, Sharda:2021}. In this paper we only want to focus on the metallicity environment, rather than its uncertain relationship with redshift evolution. 13\,Gyrs can be considered an extreme cut-off --- any binary taking longer than this time to merge will not be detectable at all. However, the higher $Z$ systems with a shorter merger time may still be observable by the gravitational wave detectors. To check this, we varied our cut-off time from 13\,Gyrs - 1\,Gyr at an interval of 1\,Gyr at every step for the $Z=0.02$ models. The variation of the median masses and spins were insignificant and none of these changed our conclusions. 

To account for GW selection effects favouring more massive binaries, we also weigh the computed chirp mass, effective and remnant spins by the effective volume (V$_\mathrm{eff}\propto$M$_\mathrm{chirp}^{5/2}$)
of the binaries, since the GW signal-to-noise ratio depends on the proximity of the merging object to the interferometer, with more massive mergers emitting louder GW signals (as discussed in \citet{Chattopadhyay:2019xye}, section 4.2).  
All further discussions in this paper will only refer to the systems in our data-set that merge in a HT, and the median values (including all plots and tables) calculated by V$_\mathrm{eff}$ weighing, rather than the intrinsic distribution, unless otherwise specified. By using the HT cut-off and biasing the mass and spin distributions by V$_\mathrm{eff}$ we account for the GW selection effects.

To investigate the spins, each model is post-processed twice under spinning and non-spinning pre-merger BH assumptions (BH-Q and BH-Z). 
For statistical robustness, we also evolve two test models with same assumptions as Pessimistic and Optimistic. 
We note that while the test models slightly change the median values, with that for Pessimistic BHNSs always $\lesssim 2.5$\% and that for their NSBH counterparts as $\lesssim 15$\%, the highest oscillation being noted for BH-Z $\chi_\mathrm{eff}$. 
However, none of these altered our final results and conclusions.

\section{Results}
\label{sec:results}

\begin{table*}
    \centering
    \begin{tabular}{llcccccccccccc}
         \multicolumn{2}{c|}{Model}&& \multicolumn{2}{c|}{$M_\mathrm{chirp}$ (M$_\odot$)}  && \multicolumn{3}{c|} {$\chi_\mathrm{rem}$}  & & \multicolumn{3}{c|}{$\chi_\mathrm{eff}$} \\
\hline
        Name & Metallicity&& BHNS & NSBH && BHNS & NSBH & NSBH&& BHNS & NSBH& NSBH\\
        &($Z$)&&&&&&(BH-Q)&(BH-Z)&&&(BH-Q)&(BH-Z)\\
\hline
Optimistic & 0.0005 && 3.32 & 4.30 && 0.42 &0.56&0.20 && e-6$^*$ & 0.41 &e-6\\
  & 0.001 && 3.65 & 3.84 && 0.39 & 0.59& 0.22 && e-6 & 0.44&e-6\\
 &0.005&&3.57&3.17 && 0.36&0.89&0.46&&e-6&0.58&e-3\\
 &0.02&& 2.29& 1.92&& 0.48&0.94&0.66&&e-6&0.59&e-6\\
\hline
Pessimistic & 0.0005 && 3.28 &2.87 && 0.44&0.94&0.67&&e-6&0.58&e-6\\
&0.001&&3.63&2.71 && 0.40&0.93&0.68&&e-6&0.55&e-6\\
&0.005&&3.58&2.26 && 0.36&0.94&0.72&&e-6&0.52&e-6\\
&0.02&&2.29&2.57 && 0.49&0.85&0.63&&e-6&0.38&e-6\\
\hline
Alpha\_2 & 0.0005 && 3.07 & 2.56 && 0.44 & 0.97&0.69&&e-6&0.62&e-6\\
&0.001&&3.25&2.60 && 0.42&0.94&0.69&&e-6&0.56&e-6\\
&0.005&&3.35&2.13 && 0.39&0.94&0.72&&e-6&0.52&e-6\\
&0.02&&2.04&2.33 && 0.56&0.75&0.64&&e-6&0.24&e-6\\
\hline
Kick\_100 & 0.0005 && 3.24&2.55 && 0.45&0.99&0.69&&e-6&0.62&e-6\\
&0.001&&3.32&2.55 && 0.42&0.95&0.69&&e-6&0.60&e-6\\
&0.005&&3.46&2.24 && 0.38&0.99&0.72&&e-6&0.58&e-6\\
&0.02&&2.28&2.17&& 0.51&1.00&0.74&&e-6&0.57&e-6\\
\hline

    \end{tabular}
    \caption{The median values of $M_\mathrm{chirp}$, $\chi_\mathrm{rem}$ and $\chi_\mathrm{eff}$ for the four models Optimistic, Pessimistic, Kick\_100 and Alpha\_2; each at four metallicities $Z=0.0005,0.001,0.005,0.02$, under two pre-merger BH spin assumptions -- BH-Q and BH-Z. The standard error (standard deviation/square root of the length of the data-set) for all BHNS models remain $<0.1$\%, the pessimistic NSBHs at $Z=0.02$ has the largest standard error $<2$\% - owing to its smaller data-set compared to the rest. $*$ $\chi_\mathrm{eff}~\mathcal{O}(-6)$ can be considered effectively zero. This table is used as a reference to compare the observations GW200105 and GW200115 in this paper and may be used similarly for future observations of NS+BH mergers.}
    \label{tab:model_median}
\end{table*}

For each model, at each metallicity, we use three well-measured observational quantities---the chirp mass ($M_\mathrm{chirp}$), effective spin ($\chi_\mathrm{eff}$) and the remnant BH spin ($\chi_\mathrm{rem}$) of NS+BH systems---and compare them to each of our models.
NSBH and BHNS systems are expected to have different mass spectrums \citep[e.g.,][]{Chattopadhyay:2020lff}, and the spin distributions of the pre-merger NS and BH of the two types of systems are also different. 
NSBHs have typically higher spins than BHNSs, since the first born NS can be spun up by accretion from the evolving companion and tidal forces can spin up the progenitor of the second born BH 
(for the BH spin model BH-Q). 
The remnant spins $\chi_\mathrm{rem}$ are also expected to be distinct between the two sub-populations of NS+BHs, even under the assumption of non-spinning BH model (BH-Z), since the mass ratios of the BHNSs are distinct from the NSBHs (see Equation~\ref{equ:chi_remnant}). 
The detailed differences between NSBH and BHNS systems, including their possible observational signatures both through mergers (gravitational waves) as well as pulsars (radio), are discussed in \citet{Chattopadhyay:2020lff}. 
As the mass and spin distributions are different, electromagnetic counterparts of NS+BH mergers such as short gamma-ray bursts can be expected to produce different signatures for the two different sub-populations \citep{Zhu:2021,Sarin:2022cmu}. 
Although, having a non-spinning pre-merger BH, reduces the chances of detecting other electromagnetic counterparts of NS+BH mergers \citep{Barbieri:2019kli}. To further, the mass ratio distribution has a higher uncertainty than $\chi_\mathrm{rem}$ and for non-spinning BHs ($\chi_\mathrm{BH}\neq0$), there is no one-to-one correspondence between mass ratio and remnant spin (see eqn.~\ref{eq:aBHfitQin},\ref{equ:chi_remnant_coefficients}). Thus $\chi_\mathrm{rem}$ proves to be a key quantity that reveals the differences between sub-populations, yielding information about the observed system's past and post-merger properties.

To analyse the results, we compare the well-measured observational values of $M_\mathrm{chirp}$, $\chi_\mathrm{eff}$ and  $\chi_\mathrm{rem}$ (along with their uncertainty ranges) to the distributions predicted by our models.
If the observed values fall within the 10$^\mathrm{th}$--90$^\mathrm{th}$ percentile predicted by our model, we declare that the model cannot be ruled out by the data, and is ``accepted". 
If the observational values lie beyond the 0.1$^\mathrm{th}$-100$^\mathrm{th}$ percentile of the simulated value, we identify the model as ``rejected" for that observation. 
When an observational data-point lies between 0.1$^\mathrm{th}$--10$^\mathrm{th}$ percentile and 90$^\mathrm{th}$--100$^\mathrm{th}$ percentile of the computed model, we tag the model as a ``potential match". 

The median value of  $M_\mathrm{chirp}$, $\chi_\mathrm{eff}$ and  $\chi_\mathrm{rem}$ from the $4\times4$ grid of models (with 4 stellar/binary evolution prescriptions and at 4 metallicities) are presented in Table~\ref{tab:model_median}. 

We compare the distribution of $M_\mathrm{chirp}$, $\chi_\mathrm{rem}$ and $\chi_\mathrm{eff}$ with the two NS+BH GW observations. 
The model distributions are shown with the 10$^\mathrm{th}$, 50$^\mathrm{th}$ and 90$^\mathrm{th}$ percentiles.
These data vs. model mass-spin comparison plots are only presented for Pessimistic and Optimistic models.
Analysis of the other two pessimistic variations -- Kick\_100 and Alpha\_2  -- yields nearly the same final conclusion as Pessimistic, although independent comparisons of the mass or either of the spin measurements may 
vary.

As mentioned in Section~\ref{sec:methods}, we assume that the spins of the NS and BH prior to merger are aligned with the orbit to calculate $\chi_\mathrm{eff}$. 
There are instances in our data-set of models accepted by $M_\mathrm{chirp}$ analysis being rejected by $\chi_\mathrm{rem}$ (e.g., NSBH at $Z=0.02$ of model Optimistic for GW200115 as shown in Fig~\ref{fig:Mchirp},  Fig~\ref{fig:chi_rem}) and vice-versa (e.g. NSBH at $Z=0.001$ of model Optimistic,BH-Z for GW200115 as shown in Fig~\ref{fig:Mchirp}, Fig~\ref{fig:chi_rem}). 
It is hence concluded that at least under the aligned spin assumption, $\chi_\mathrm{rem}$ and $M_\mathrm{chirp}$ are necessary and sufficient in determining the feasibility of a model. 
Using only $\chi_\mathrm{eff}$ and $M_\mathrm{chirp}$ as pointers may lead to wrong conclusions.

\subsection{Chirp mass}
\label{subsec:Mc}

\begin{figure}
    \centering
    \includegraphics[width=\columnwidth]{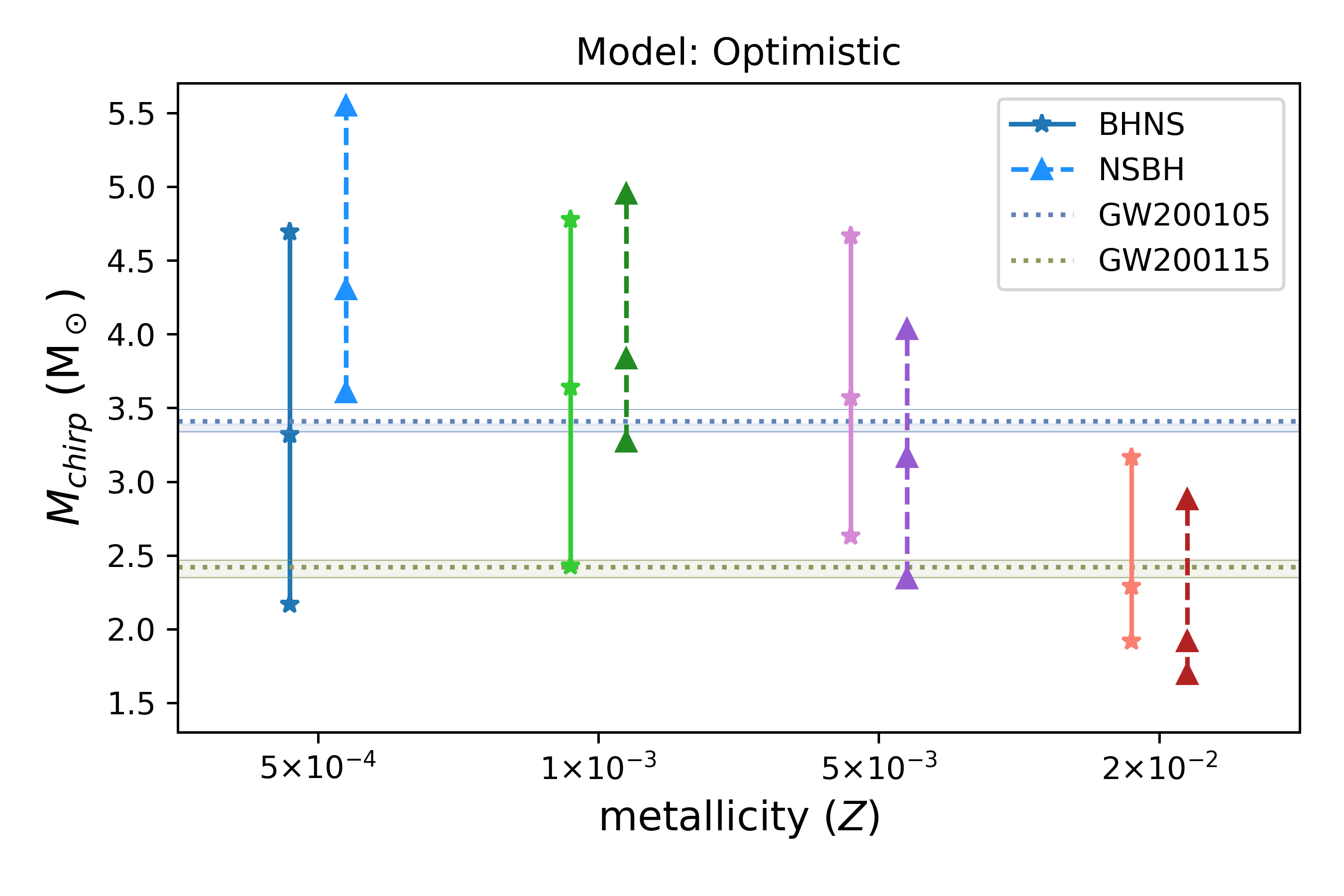}
    \caption{Distribution of chirp masses $M_\mathrm{chirp}$ of BHNS and NSBH binaries (10$^\mathrm{th}$ percentile, median and 90$^\mathrm{th}$ percentile range) as a function of metallicity for model Optimistic. The median values are non-intrinsic, i.e. only consists of systems that merge in a Hubble time, weighted by the effective volume V$_\mathrm{eff}$.
    As described in the paper, the sample is obtained for the net population evolved that merges in a HT, weighted by their respective $V_\mathrm{eff}$ to account for selection biases. 
    The BHNSs are shown with star markers and solid lines, while the NSBHs can be identified by triangle markers and dashed lines. 
    The shaded horizontal lines show the $M_\mathrm{chirp}$ for GW200105 and GW200115 \citep{LIGOScientific:2021qlt}. 
    }
    \label{fig:Mchirp}
\end{figure}

\begin{figure}
    \centering
    \includegraphics[width=\columnwidth]{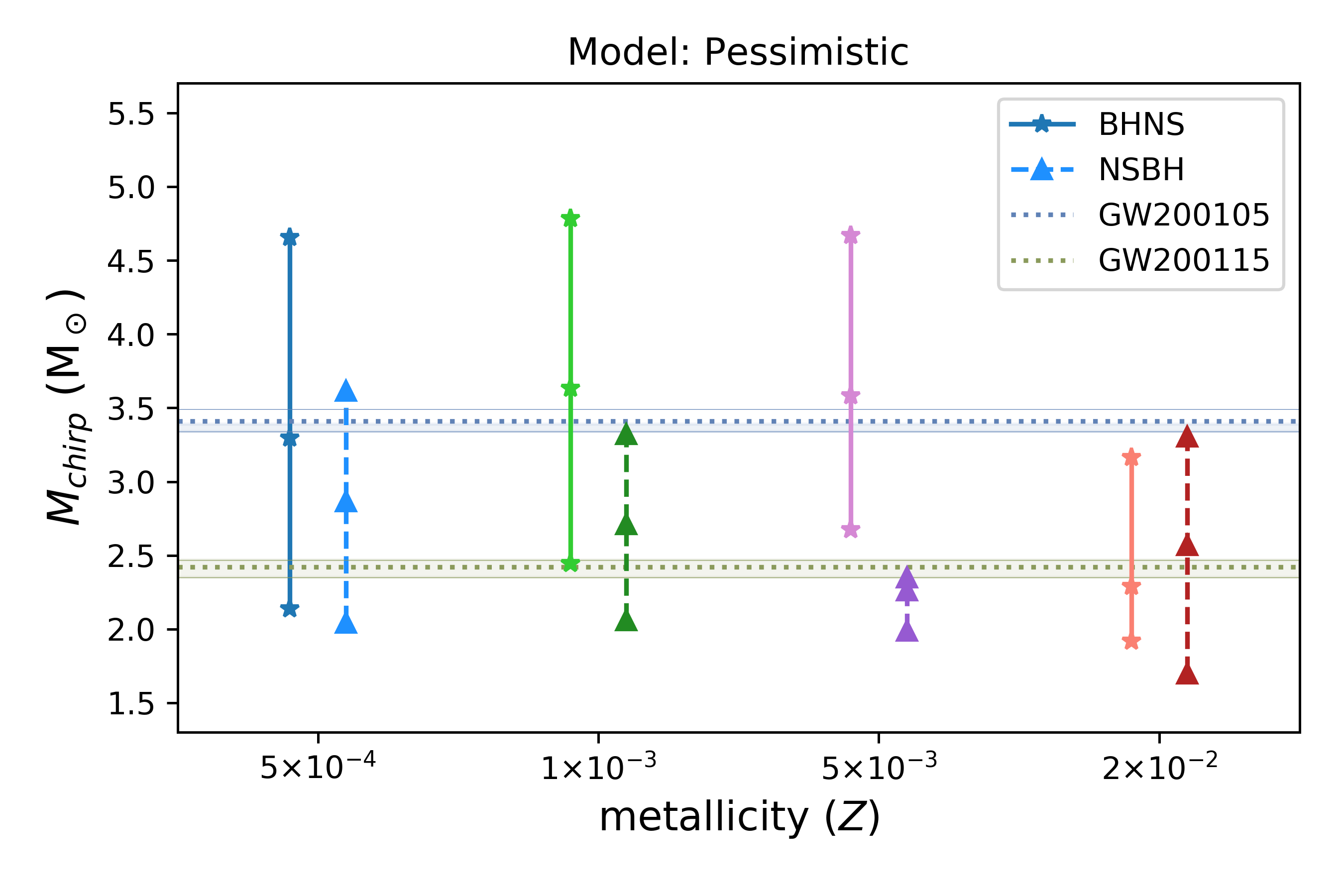}
    \caption{Chirp mass distribution for model Pessimistic with markers identical as that of Fig.~\ref{fig:Mchirp}. 
    }
    \label{fig:Mchirp_pess}
\end{figure}

The $M_\mathrm{chirp}$ variation in models Pessimistic and Optimistic is shown in Fig.~\ref{fig:Mchirp} and Fig.~\ref{fig:Mchirp_pess} respectively. 

In the model Pessimistic, the BHNS chirp mass median always appear to be slightly larger than NSBHs, while for Optimistic, this gets reversed at lower metallicities. 
This appears paradoxical, as the dominant formation channels described in \citet{Chattopadhyay:2020lff} for solar metallicity shows the BHs of BHNSs to be more massive - as the more massive star evolves faster and remains more massive enough to form BHs after mass transfer. 
For NSBHs in this case, the ZAMSs masses of the NS and BH progenitors are nearly equal - where the slightly larger mass star transfers enough mass to its companion to form a NS, whereas the slower evolving companion becomes a BH. 
However, this formation story changes for lower metallicities. 
For one, reduced stellar winds ensure the stars lose less mass - creating more massive BHs that would merge in a shorter timescale than their less massive counterparts \citep{Peters:1964}. 
Secondly, stellar winds may also increase orbital separation. Hence, in metal-poor stars, lower wind loss is also less likely to decrease the orbital separation of the binary - lower orbital separation being crucial for forming NSBHs (as well as BHNSs that merge within a HT).
This ensures that while BHNS for model Optimistic at $Z = 0.0005$ form from ZAMS stars with orbital separation roughly 0.53\,AU, NSBHs form from binaries with initial separation of about 0.2 of that. 
The median ZAMSs masses of the two sub-populations NSBH and BHNSs at Optimistic $Z=0.0005$ are (m$_\mathrm{BH}^\mathrm{ZAMS} = 26.58$\,M$_\odot$, m$_\mathrm{NS}^\mathrm{ZAMS} = 32.65$\,M$_\odot$) and (m$_\mathrm{BH}^\mathrm{ZAMS} = 33.64$\,M$_\odot$, m$_\mathrm{NS}^\mathrm{ZAMS} = 19.19$\,M$_\odot$). 
At low metallicity, the NSBH progenitor-NS (which remain more massive due to less wind loss), due to close proximity with its companion, starts unstable mass transfer ensuring a CE phase, donating even more mass to its companion. 
The companion being already massive and also losing less mass for reduced stellar winds, becomes a BH. 
The BHNSs form from systems with a more unequal mass ratio and larger orbital separation, and an HG donor CE phase is not as essential in its formation as for a NSBH. 
Hence we obtain the double compact median masses of NSBHs and BHNSs as {$M_\mathrm{BH} = 20.67$\,M$_\odot$, $M_\mathrm{NS} = 1.50$\,M$_\odot$} and {$M_\mathrm{BH} = 5.64$\,M$_\odot$, $M_\mathrm{NS} = 1.56$\,M$_\odot$} respectively. 
It is indeed interesting to note that the masses of NSBHs are so asymmetric at low metallicity under Optimistic assumption. 
To compare, we find at $Z = 0.02$, the median masses of the NSBHs and BHNSs are ($M_\mathrm{BH} = 3.11$\,M$_\odot$, $M_\mathrm{NS} = 1.46$\,M$_\odot$)
and ($M_\mathrm{BH} = 3.14$\,M$_\odot$, $M_\mathrm{NS} = 1.37$\,M$_\odot$) respectively. 
In the Pessimistic scenario, as the HG donor star does not survive the CE, this specific channel---which is dominant for Optimistic NSBHs---is suppressed, and the formation follows the more regular pathway of mass reversal at a higher ZAMS separation. 
See \citet{Broekgaarden:2020NSBH} for more discussion on different NS+BH formation channels and \citet{Chattopadhyay:2020lff} for more discussion on the mass, spin and orbital properties of NS+BH at solar and sub-solar metallicities.  

The $M_\mathrm{chirp}$ of GW200105 is higher than GW200115. 
For Pessimistic, both GW200105 and GW200115 are in agreement with BHNSs and NSBHs at $Z=0.0005,0.001$, while at $Z=0.005$ GW200105 matches with the BHNS and GW200115 with the NSBH sub-populations. 
At the higher metallicity of $Z=0.02$, only the less massive GW200115 can remain consistent with the BHNSs and NSBHs. 
The metal-rich Pessimistic NSBHs, however, also show a potential match with GW200105. 
In the Optimistic scenario, GW200105 matches with both BHNS and NSBH systems at $Z=0.001,0.005$ and BHNSs at $Z=0.0005$. 
NSBHs at $Z=0.0005$ rests in a potential match zone for GW200105. 
Only at the very lower limit, does GW200115 match with the BHNSs of Optimistic at $Z=0.0005,0.001,0.005$ and NSBH at $Z=0.001$. 
At the high $Z$ end of 0.02, only GW200115 shows good agreement with both BHNSs and NSBHs. 

The median $M_\mathrm{chirp}$ of GW200105 is 3.41\,M$_\odot$,  which matches well with the $M_\mathrm{chirp}$ of BHNS for all of our models with $Z\leq0.005$. 
The less massive GW200115, with an $M_\mathrm{chirp}$ of about 2.42\,M$_\odot$, is a better match with the median NSBHs of all three pessimistic group of models at all $Z$, and with BHNSs of all four models at $Z=0.02$.

Models Alpha\_2 and Kick\_100 produces identical consequences in terms of $M_\mathrm{chirp}$.

\subsection{Remnant spin}
\label{subsec:chirem}

\begin{figure}
    \centering
    \includegraphics[width=0.9\columnwidth]{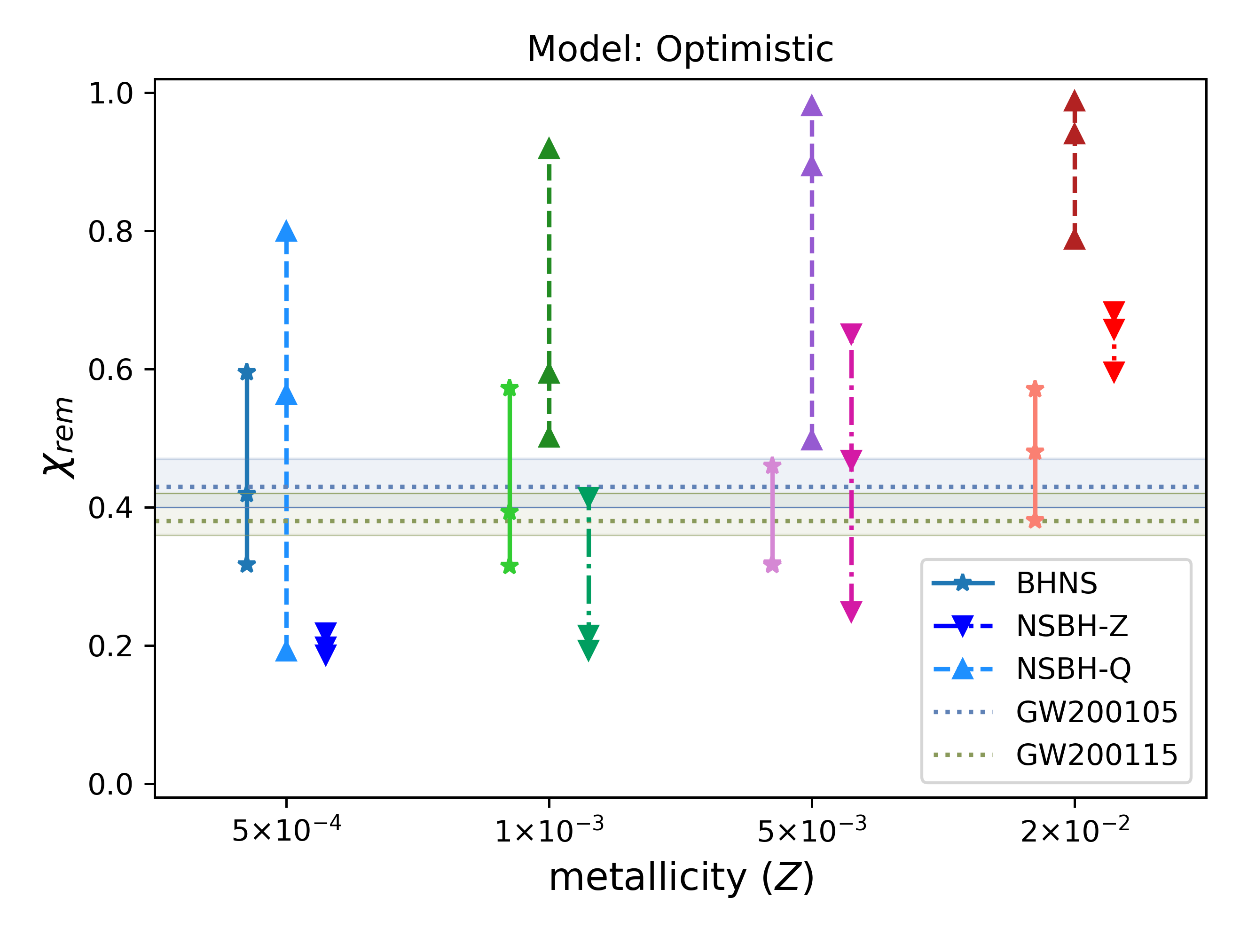}
    \caption{Distribution of the spin of BHNS and NSBH merger remnants $\chi_\mathrm{rem}$ (at 10$^\mathrm{th}$ percentile, median and 90$^\mathrm{th}$ percentile range) as a function of metallicity. The BH-Z NSBHs are plotted with inverted triangles (and dash-dotted lines) and the BH-Q NSBHs with upright triangles (and dashed lines) as markers. The BHNSs, identical for BH-Q and BH-Z are shown in the symbol of stars (and solid lines). The shaded horizontal regions show the inferred values of GW200105 and GW200115 \citep{LIGOScientific:2021qlt}. For additional details see the main text.}
    \label{fig:chi_rem}
\end{figure}

\begin{figure}
\centering
    \includegraphics[width=0.9\columnwidth]{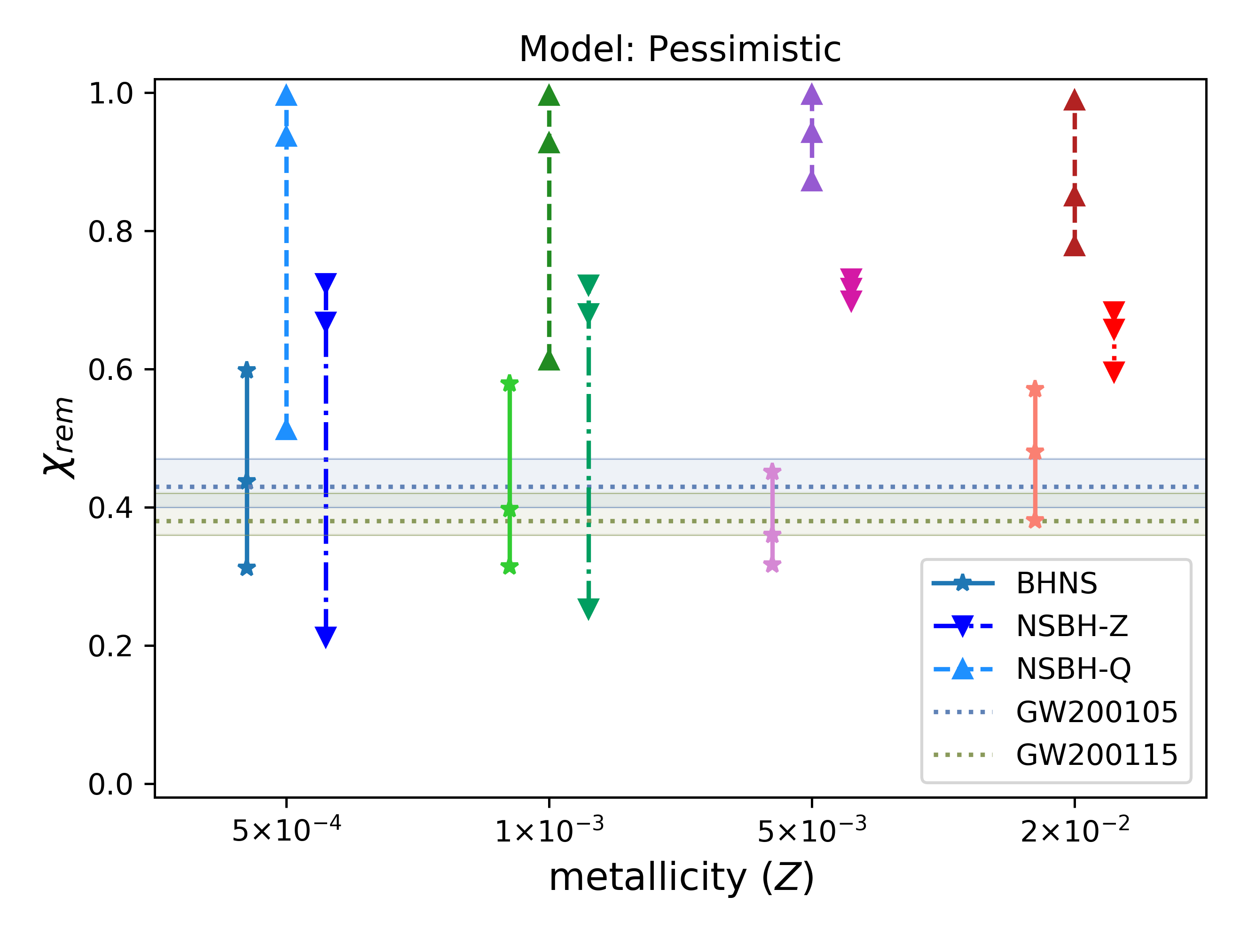}
    \caption{The remnant spin distribution of model Pessimistic at four metallicities. The markers are identical as Fig.~\ref{fig:chi_rem}.}
    \label{fig:chi_rem_pess}
\end{figure}

Fig.~\ref{fig:chi_rem} and ~\ref{fig:chi_rem_pess} shows the $\chi_\mathrm{rem}$ distribution for each metallicity grid under spinning (BH-Q) and non-spinning (BH-Z) assumptions for models Optimistic and Pessimistic, respectively. 
It is clear that the predictions for NSBHs strongly vary between the BH-Q and BH-Z assumptions.
BHNSs always have a non-spinning pre-merger BH and hence have the same $\chi_\mathrm{rem}$ distribution for both the assumptions. 

For Optimistic (Fig.~\ref{fig:chi_rem}), we observe that BHNSs at all metallicities are permissible for both GW200105 and GW200115. 
BH-Q NSBHs at $Z=0.0005$ and BH-Z NSBHs at intermediate metallicities $Z=0.001,0.005$ also match with both the observations. 
BH-Q NSBHs at $Z=0.001,0.005$ lies on the boundary as potential matches for GW200105. 

Equation~\ref{equ:chi_remnant} and equation~\ref{equ:chi_remnant_coefficients} shows that even for non-spinning ($\chi_\mathrm{BH} = 0$) pre-merger BHs, the remnant BH will be spinning, since the term $\gamma$ and its coefficients remain non-zero. 
Even for BH-Z, $\chi_\mathrm{rem} \gtrsim 0.15$ for Optimistic.  

As $M_\mathrm{BH}$>$M_\mathrm{NS}$, more asymmetric mass ratio (effectively, more massive pre-merger BHs) ensure $\gamma$ is small, making $\chi_\mathrm{rem}$ small. As described in section.~\ref{subsec:Mc}, metal-poor NSBHs have fairly massive BHs. This, with the BH-Z assumption, makes NSBH $\chi_\mathrm{rem}$ at $Z=0.0005,0.001$ of Optimistic quite modest. For BH-Q, even with highly asymmetric mass ratio, these low metallicity NSBHs show median $\chi_\mathrm{rem}\approx0.55$. Guided by equation~\ref{eq:aBHfitQin}, small orbital separation (and hence shorter period $p_\mathrm{orb}$) between the two stars in these systems (as discussed in section.~\ref{subsec:Mc}) aids in producing 
higher $\chi_\mathrm{BH}$. 

In the model Pessimistic (Fig.~\ref{fig:chi_rem_pess}), BHNSs at all metallicities match with both of the GW observations. Even with large median values of $\chi_\mathrm{rem}$, BH-Z NSBHs at $Z=0.0005,0.001$ are consistent with GW200105 and GW200115. 
The only difference between Pessimistic and Kick\_100, Alpha\_2 is observed for $Z=0.02$ $\chi_\mathrm{rem}$ for BHNSs. While for Pessimistic, this is permissible for GW200115, both Kick\_100 and Alpha\_2 discludes these variations. 

Solely observing the $\chi_\mathrm{rem}$ medians in Table.~\ref{tab:model_median}, and comparing that with the observations of GW200105 and GW200115 with remnant spins of about 0.43 and 0.38 respectively, we see that apart from metal-poor Optimistic models, all other models with BH-Q spin assumptions have quite large $\chi_\mathrm{rem}$ medians. All BHNSs in models with $Z\leq0.005$ match fairly well with the two observations. However, BH-Z NSBHs, though still having higher median, show some matches (see  Fig.~\ref{fig:chi_rem} and \ref{fig:chi_rem_pess}). 

We also check the $\chi_\mathrm{rem}$ PessimisticST model, replacing $\chi_\mathrm{BH}$ of eqn.~\ref{equ:chi_remnant_coefficients} with $\chi_\mathrm{BH}\cos{\theta}$ (where $\theta$ is the spin misalignment angle of the pre-merger BH with the orbital angular momentum). The computed $\chi_\mathrm{rem}$ does not alter our conclusions, only bringing the NSBH-Q to slightly ($<$0.1) lower values, making $Z=0.0005$ NSBH-Q $\chi_\mathrm{rem}$ a plausible model for both GW200105 and GW200115 for its 10$^\mathrm{th}$ percentile boundary (the median value $\approx 0.75$ still remains well above the observations). Since the first born BH of the BHNSs are always assumed non-spinning, we further compute a special variation of this population with the magnitude of the first born BH spin $\chi_\mathrm{BH}=0.1$, making $\chi_\mathrm{BH}\cos{\theta}$ non-zero for BHNSs. Still, this does not alter our conclusions and the $\chi_\mathrm{rem}$ distribution is very slightly adjusted to smaller values \footnote{It allows $Z=0.0005$ BHNS to have the 10$^\mathrm{th}$ percentile of $\chi_\mathrm{rem} = -0.001$, but all changes for other models are unremarkable. We hence see that at low metallicity, anti-aligned merger remnant is possible due to pre-merger BH orbital mis-alignment, but the magnitude of the $\chi_\mathrm{rem}$ still remains very close to zero.}

\subsection{Effective spin}
\label{subsec:chieff}

\begin{figure}
    \centering
    \includegraphics[width=0.9\columnwidth]{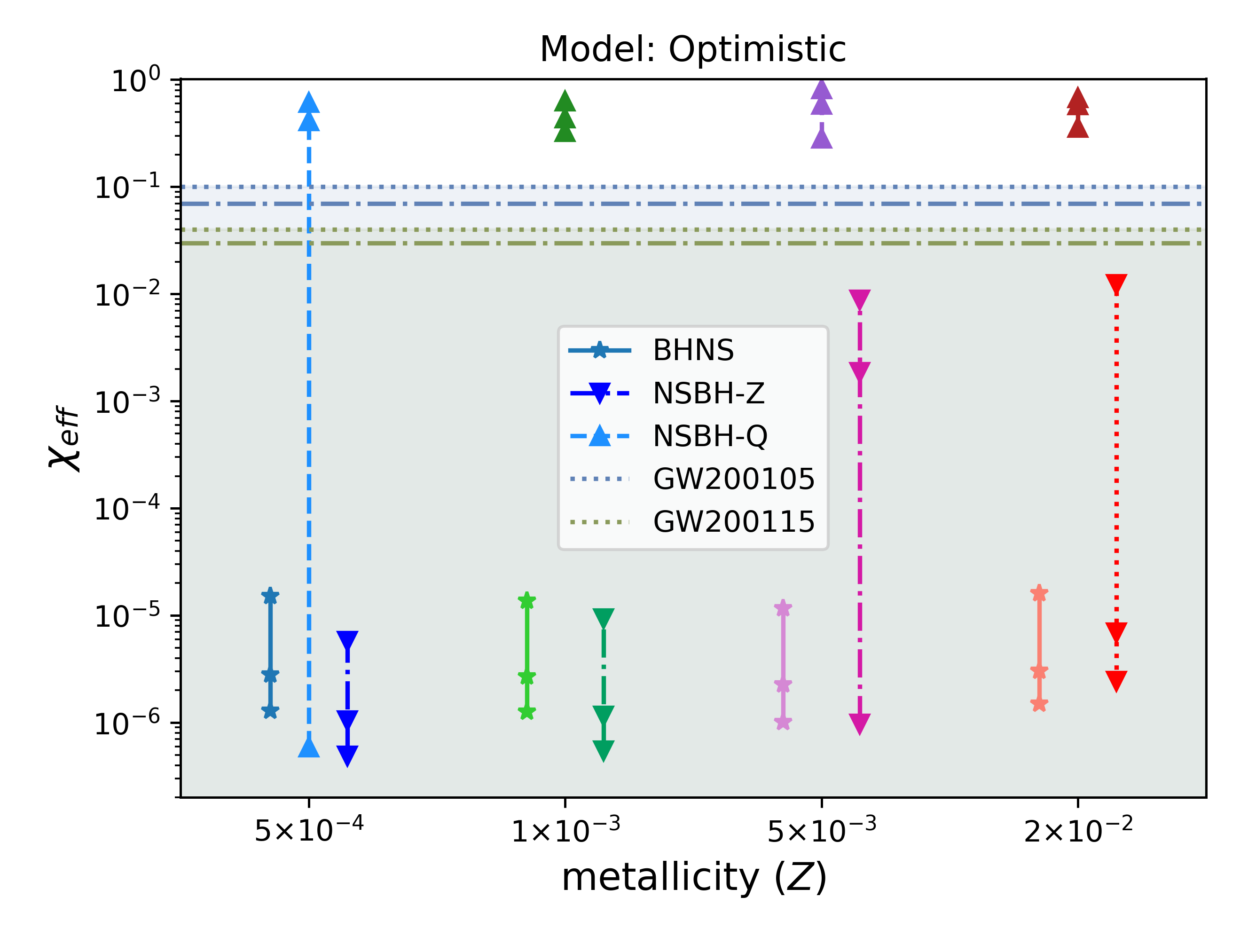}
    \caption{Effective spin distribution of model Optimistic with markers identical to Fig.~\ref{fig:chi_rem}. The horizontal lines in blue and green show the upper limits of $\chi_\mathrm{eff}$ observations of GW200105 and GW200115, respectively.
    Of these horizontal lines, the dotted and dash-dotted lines are the upper limits for a higher and lower spin priors respectively. Since we only assume aligned spin model, we show the part of observations (shaded region), where $\chi_\mathrm{eff}>0$.   }
    \label{fig:chi_eff_opti}
\end{figure}

\begin{figure}
    \centering
    \includegraphics[width=0.9\columnwidth]{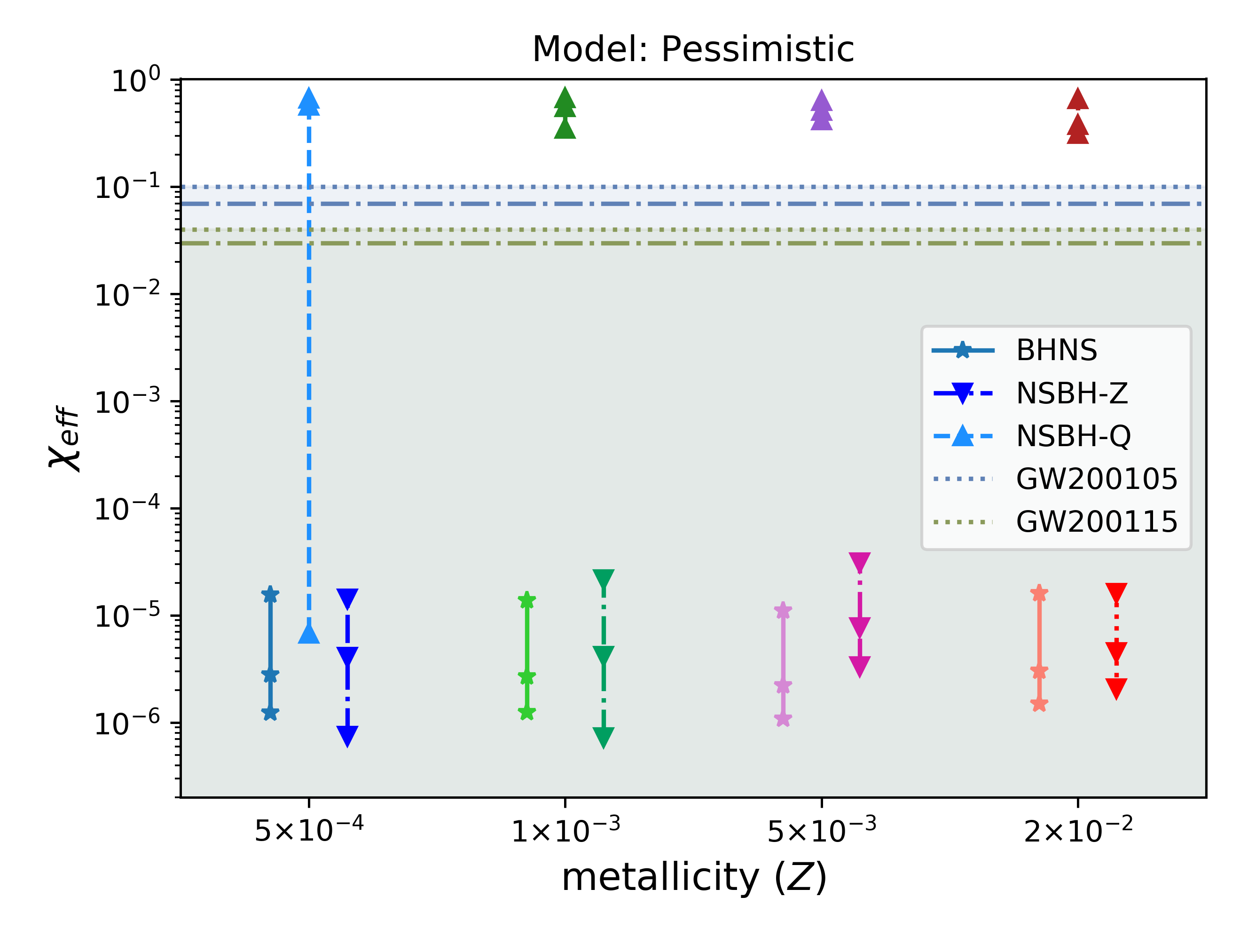}
    \caption{Effective spin distribution for model Pessimistic with identical symbols as Fig.~\ref{fig:chi_eff_opti}.}
    \label{fig:chi_eff_pess}
\end{figure}

Although we have described that $M_\mathrm{chirp}$ and $\chi_\mathrm{rem}$ are together sufficient in constraining our models, we show the distribution of $\chi_\mathrm{eff}$ for models Optimistic and Pessimistic (for both spin models BH-Q and BH-Z) in Fig.~\ref{fig:chi_eff_opti} and Fig.~\ref{fig:chi_eff_pess} respectively.

We note here that for $\chi_\mathrm{eff}$, both Optimistic and Pessimistic  (and hence all four models) lead to identical conclusions. We observe that BHNSs at all of the metallicities modelled, as well as all NSBHs with BH-Z assumption and metal poor ($Z=0.0005$) BH-Q NSBHs are congruent with both the GW mergers. The large uncertainty of $\chi_\mathrm{eff}$ indeed puts poor constraints for the models.

\begin{figure}
    \centering
    \includegraphics[width=0.9\columnwidth]{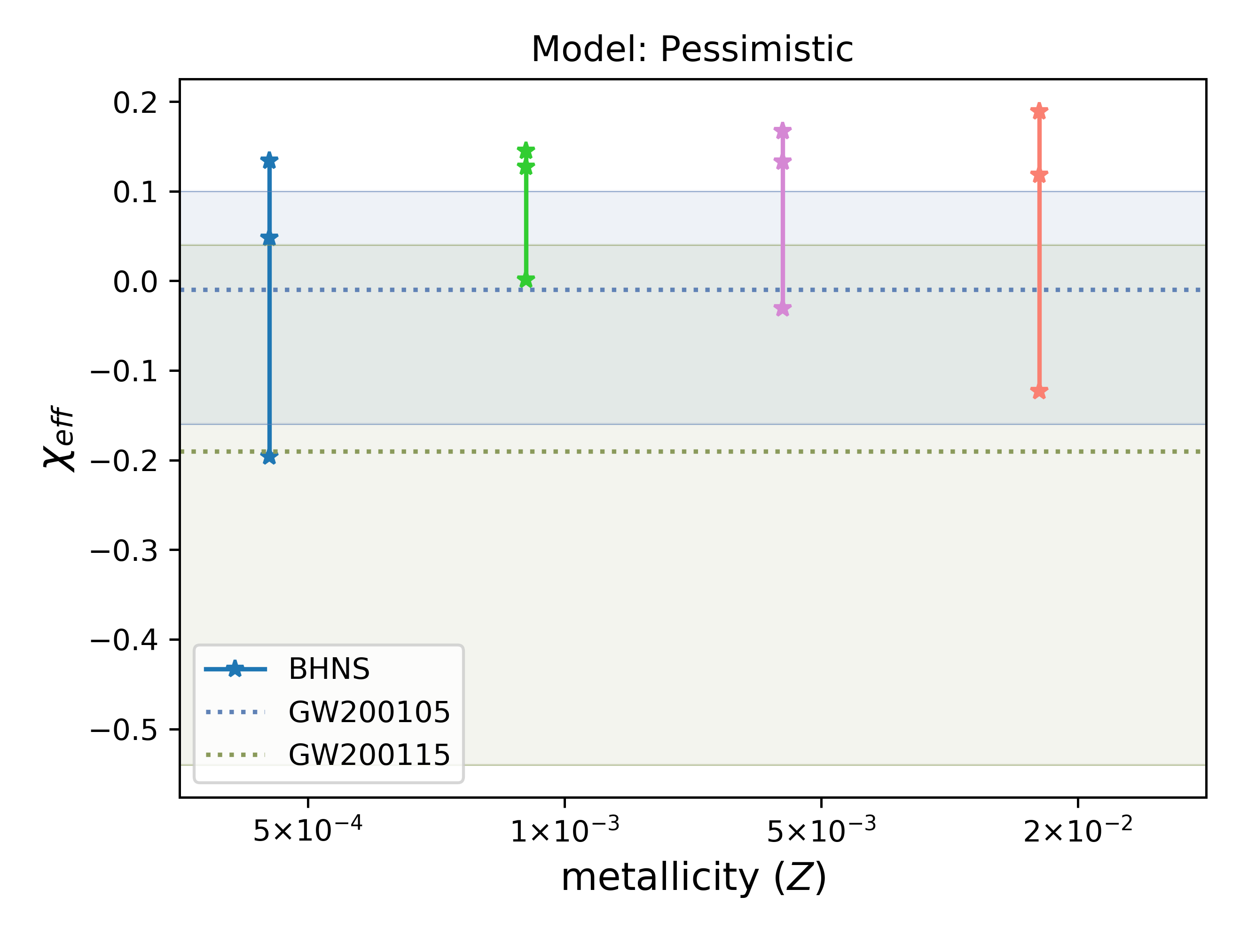}
    \caption{Effective spin distribution for model PessimisticST, showing the BHNSs. The shaded region for GW200105 and GW200115 are as inferred by \citet{LIGOScientific:2021qlt}, under high secondary spin prior.We note that although spin mis-alignment allows for negative $\chi_\mathrm{eff}$, the median of all distributions are greater than 0.}
    \label{fig:ST_effSpin}
\end{figure}

The primary contribution to $\chi_\mathrm{eff}$ comes from the spin of the BH, owing to its higher mass than the NS (see equation ~\ref{eq:chi_effec}). 
Hence, fast spinning BHs of BH-Q can impart NSBHs with much higher $\chi_\mathrm{eff}$ than BHNS. 
For BH-Z on the other hand, NSBHs consisting of very massive, non-spinning BHs tend to have larger denominators in equation ~\ref{eq:chi_effec}, resulting in small $\chi_\mathrm{eff}$ values (as seen in Optimistic BH-Z NSBHs at $Z=0.0005,0.001$). 
The PessimisticST model that allows for spin misalignment is still shown to be biased towards zero or positive $\chi_\mathrm{eff}$, allowing negative values only in a minority of the systems. 
We note that PessimisticST again, does not alter our conclusions. 

\subsection{Constraints from mass and spin}
\label{subsec:constraints}

\begin{table}
    \centering
    \begin{tabular}{llccc}
    \multicolumn{2}{c|}{Model}& \multicolumn{3}{c|}{Results}\\
    \hline
    Name & Metallicity (Z) & \multicolumn{2}{c|}{NSBH} & BHNS\\
    -&-&(BH-Q)&(BH-Z)&\\
    \hline
    Optimistic & 0.0005 &gw05$^\dagger$  &-& gw05,gw15\\
     & 0.001 &gw05$^\dagger$ &gw05& gw05,gw15\\
     & 0.005 &gw05$^\dagger$ &gw05,gw15& gw05\\
     & 0.02 & - &-&gw15 \\
     \hline
    Pessimistic & 0.0005 & -&gw05,gw15& gw05,gw15\\
     & 0.001 &- &gw05$^\dagger$,gw15& gw05,gw15\\
     & 0.005 &- &-& gw05,gw15 \\
     & 0.02 & -&-& gw15\\
     \hline
    Alpha\_2 & 0.0005 & -&gw05,gw15& gw05,gw15\\
     & 0.001 &- &gw05$^\dagger$,gw15& gw05,gw15\\
     & 0.005 &- &-& gw05,gw15 \\
     & 0.02 & -&-& -\\
     \hline
    Kick\_100 & 0.0005 & -&gw05,gw15& gw05,gw15\\
     & 0.001 &- &gw05$^\dagger$,gw15& gw05,gw15\\
     & 0.005 &- &-& gw05,gw15 \\
     & 0.02 & -&-& -\\
     \hline
    \end{tabular}
    \caption{The grid of models and their match with the two NS+BH merger observations through mass and spin analysis. The models that are \emph{accepted} are denoted by `gw05' or `gw15' for GW200105 or GW200115 respectively. The \emph{potential match} models are identified by a dagger($\dagger$) symbol and the \emph{rejected} ones are dashed(-). }
    \label{tab:results}
\end{table}

Using the analysis of section.~\ref{subsec:Mc}, ~\ref{subsec:chirem} and ~\ref{subsec:chieff}, we produce the following conclusions from each of the four models in the Table~\ref{tab:results}. The summary of the Table~\ref{tab:results} is given here ---

\begin{enumerate}
\item Optimistic: \\
GW200105--- \emph{accepted} for BHNS at $Z=0.0005,0.001,0.005$, NSBH BH-Z at $Z=0.001,0.005$; \emph{potential match} NSBH BH-Q at $Z=0.0005,0.001,0.005$, NSBH BH-Z at $Z=0.0005$; \emph{rejected} BHNS or NSBH at $Z-0.02$.\\
GW200115--- \emph{accepted} BHNS at $Z=0.0005,0.001$, BHNS at $Z=0.02$, NSBH BH-Z at $Z=0.005$;  \emph{rejected} for BHNS at $Z=0.005$.\\
\item Pessimistic, Alpha\_2, Kick\_100: \\
GW200105--- \emph{accepted} for BHNS at $Z=0.0005,0.001,0.005$, NSBH BH-Z at $Z=0.0005$, \emph{potential match} for NSBH BH-Z at $Z=0.001$.\\
GW200115--- \emph{accepted} for BHNS at $Z=0.0005,0.001,0.005$, NSBH BH-Z at $Z=0.0005,0.001$.
GW200115 is \emph{accepted} for BHNS at $Z=0.02$ for Pessimistic, \emph{rejected} for Alpha\_2, Kick\_100. 

It is to be noted that only Alpha\_2 and Kick\_100 are the models that matches with the results of \citet{Broekgaarden:2021hlu}.

\end{enumerate}

We note that the qualitative conclusions from the Optimistic and Pessimistic models are not vastly different from each other. 
Sub-solar metallicity and BHNSs are preferred. 
For Alpha\_2 and Kick\_100, no $Z=0.02$ population are consistent with the two GW observations. 
Only with GW200115 does metal-rich $Z = 0.02$ Optimistic BHNSs show a consistency. 
The NSBHs that are in agreement with observations for all four models are for non-spinning pre-merger BH (i.e. spin model BH-Z). 
Only for Optimistic does GW200105 shows a potential matching with BH-Q NSBHs.  

Emphasizing the median values of Table~\ref{tab:model_median}, $Z\lesssim0.005$ BHNSs show a better match of $M_\mathrm{chirp}$ and $\chi_\mathrm{rem}$ of GW200105. 
The less massive GW200115 fits better with metal rich values of $M_\mathrm{chirp}$ for BHNSs of all models, and all NSBHs values of the pessimistic group of models. 
The $\chi_\mathrm{rem}$ medians of NSBHs are however too high for GW200115 NSBHs (especially BH-Q models). 
This makes either a metal-rich BHNS or BH-Z NSBH a more likely source of GW200115.

\subsection{Relative merger rate}
\label{subsec:rates}

\begin{table}
    \centering
    \begin{tabular}{llcc}
    \multicolumn{2}{c|}{Model}& \multicolumn{2}{c|}{Rates}\\
    \hline
    Name & Metallicity (Z) & $\mathcal{R_\mathrm{NSBH/BHNS}}$ & $\mathcal{R_\mathrm{norm. BHNS}}$\\
    \hline
    Optimistic & 0.0005 & 1.79e-1 & 1\\
     & 0.001 & 2.91e-1& 1.07\\
     & 0.005 &1.94e-1 & 2.17\\
     & 0.02 & 6.89e-2 & 0.43\\
     \hline
    Pessimistic & 0.0005 & 6.37e-2& 0.88\\
     & 0.001 & 9.59e-2 & 0.94\\
     & 0.005 & 5.59e-3 & 1.99\\
     & 0.02 & 3.77e-3 & 0.29\\
     \hline
    Alpha\_2 & 0.0005 &9.00e-2 & 2.59\\
     & 0.001 & 8.89e-2& 2.30\\
     & 0.005 & 3.23e-2& 3.10\\
     & 0.02 & 3.84e-3& 0.34\\
     \hline
   Kick\_100 & 0.0005 & 9.94e-2& 2.29\\
     & 0.001 & 0.57e-2& 2.37\\
     & 0.005 & 9.91e-3& 4.71\\
     & 0.02 & 9.88e-4& 1.64\\
     \hline
    \end{tabular}
    \caption{The relative merger rates of NS+BH systems for our grid of models. Where, $\mathcal{R_\mathrm{NSBH/BHNS}}=\mathrm{NSBH }/\mathrm{BHNS}$ that merges in a HT for that model and $\mathcal{R_\mathrm{norm. BHNS}}=\mathrm{BHNS }/(\mathrm{BHNS, Optimistic, Z=0.0005}$), that merges in a HT.}
    \label{tab:rates}
\end{table}

The relative merger rates of NS+BH systems are presented in Table~\ref{tab:rates}, which only accounts for the systems that merges in a HT. 
The column $\mathcal{R_\mathrm{NSBH/BHNS}}$ shows the ratio of merging NSBHs to merging BHNSs in each model. 
While all models produce more BHNSs, Optimistic $Z=0.001$ produces the largest fraction of NSBHs. 
The column $\mathcal{R_\mathrm{norm. BHNS}}$ marks the merging BHNSs in each model normalized by the merging BHNSs in Optimistic $Z=0.0005$. 
We observe that Kick\_100 at $Z = 0.005$ has the most BHNSs. It is further noted that since BHNS dominates the population significantly at all metallicities - hence non-spinning pre-merger BHs being more common - the chances of having an electromagnetic counterpart for NS+BH mergers is rare \citep{ Barbieri:2019kli,Hu:2022ubh}.
For the cosmological merger rates, we refer to \citet{Broekgaarden:2020NSBH,Broekgaarden:2021hlu}. 
We note that the cosmological rates are in agreement with our model, but large uncertainties in predictions \citep{Neijssel:2019,Broekgaarden:2021hlu}. 
The LIGO/Virgo inferred local merger rate medians vary (12--242\,Gpc$^{-3}$yr$^{-1}$) depending on assumptions made about the distribution of the individual component masses, and have broad lower/upper limits \citep{LIGOScientific:2021qlt}.  

\section{Conclusions}
\label{sec:conclusions}

We analysed the first two observations of NS+BH systems by LIGO/Virgo \citep{LIGOScientific:2021qlt} using the rapid binary population synthesis code COMPAS \citep{Stevenson:2017tfq,Vigna-Gomez:2018dza}. 
We compared these observations to four different models with various different assumptions regarding poorly constrained stages of massive binary evolution (the common envelope phase and natal kicks obtained in supernovae). 
For each model, we examined how the properties of NS+BH vary with metallicity, using a grid of four 
different metallicities ($Z=0.0005, 0.001, 0.005, 0.02$). 
We used several well measured observables (the binary chirp mass, the effective spin and the spin of the remnant black hole) as constraints on our models and showed that under the assumption of aligned spins, the observations are well in agreement with isolated binary evolution. 

The observational data for GW200115 shows possible evidence for significant spin-orbit misalignment  \citep{LIGOScientific:2021qlt}. 
Several authors have argued that high natal kicks would be necessary to mis-align the spin to such a high degree \citep{Fragione:2021qtg,Gompertz:2021xub,Zhu:2021zxj}.
However, \citet{Broekgaarden:2021hlu} showed smaller supernova kicks provide a better match with observations once the rates of all compact object mergers are taken into account.
Moreover, it has been shown that the spin alignment results of the double BH observations are quite prior-dependant \citep{Galaudage:2021}, and even for the two NS+BH mergers, a different choice yields non-spinning pre-merger BHs, not requiring any mis-alignment \citep{2021:Ilya}. 
We show by model PessimisticST that although misalignment by supernovae kick may explain the negative $\chi_\mathrm{eff}$, the dominant population still have aligned, nearly zero BH spins.
We assumed aligned spins for pre-merger BHs. 
The BH spins are computed in two models of spinning BH-Q \citep{Qin:2018nuz} and non-spinning BH-Z. 
The NS spin, used for calculating the effective spin, is also modelled in detail \citep{Chattopadhyay:2020lff, Chattopadhyay:2019xye}. 
The remnant spin of the post-merger BH is calculated using fits from \citet{Zappa:2019ntl} at different metallicities.

The answers to our questions at the beginning (section~\ref{sec:intro}) can be consolidated from section~\ref{subsec:constraints} as:

a) Making use of multiple well-measured observables, such as $M_\mathrm{chirp}$ and $\chi_\mathrm{rem}$ does indeed aid in distinguishing qualitatively between two sub-populations (NSBH v BHNS). Both observations are in favour of BHNSs, as well as a few matches for NSBHs (section~\ref{subsec:constraints}).

b) In our models we find that systems in which the neutron star forms first (NSBHs) typically contribute $\lesssim1$\% of merging NS+BH binaries, though it varies from model to model, the range being $\sim0.1-30$\% (Table.~\ref{tab:rates}). This significantly shifts the quantitative analysis in favour of BHNSs, especially for the pessimistic group of models.
The relative merger rates of the two sub-populations BHNSs and NSBHs, as well as the relative merger rates of BHNSs in our 16 models is presented in Table.~\ref{tab:rates}. Although there are more matches with BHNSs at different metallicities for both models, with non-spinning pre-merger BH model (BH-Z), NSBHs are also in agreement with the observations. 

c) Our models show that a non-spinning black hole (prior to merger) is highly preferred for both GW200105 and GW200115 \citep[see also][]{2021:Ilya}. 
There is only one \emph{potential match} for a pre-merger spinning BH for GW200105 with model Optimistic. All three pessimistic models - Pessimistic, Alpha\_2, Kick\_100 have only non-spinning model as \emph{accepted}.

d) In our analysis, we find that the effective spin parameter $\chi_\mathrm{eff}$ does not add any extra constraints on our models.

e) Formation in a sub-solar metallicity environment ($Z\lesssim0.005$ is strongly preferred for both GW200105 and GW200115, coming from analysing the $\chi_\mathrm{rem}$ of Kick\_100 and Alpha\_2.  
However, GW200115 does match with only two metal-rich models  ($Z = 0.02$, Optimistic, Pessimistic) . 

We further note that $\chi_\mathrm{rem}$, being a well measured parameter (than, say, mass-ratio), can be utilized to differentiate between the  sub-populations BHNSs and NSBHs (BH-Q and BH-Z).

Further, the paucity of NSBHs compared to BHNSs by several orders of magnitude and the general favouring of non-spinning BH models reveals that electromagnetic counterparts to NS+BH mergers are likely quite infrequent.

Our results are solely valid for isolated binary evolution. It is also helpful to bear in mind here, that the expected merger rate of NS+BH systems through the dynamical channel of young star clusters is found plausible in some studies \citep{Rastello:2020sru}, but others find this type of binary merger to be quite low for nuclear, globular, as well as young clusters \citep{Clausen:2013GC,Petrovich_Antonini:2017,Fragione:2020ApJF}.
The uncertainties of initial conditions (e.g. mass ratio and orbital period distributions) and binary evolution processes (e.g. common envelope, supernovae kicks, post supernova core mass to compact object mass mapping) also adds to the complexities \citep{Stevenson:2017tfq, Vigna-Gomez:2018dza, Broekgaarden:2020NSBH, Chattopadhyay:2020lff}. While we did vary some of these parameters of massive binary evolution, others were kept constant. The models selected for this study were primarily guided by the models already short-listed by \citet{Broekgaarden:2021hlu} based on the chirp mass of the two NS+BH mergers and the merger rates of all types of observed LIGO/Virgo mergers from the first two observing runs (Alpha\_2 and Kick\_100). 
Two other models Pessimistic and Optimistic were also added to follow the effect of this change in parameters on our results.
We acknowledge that while a portion of the partially constrained massive binary parameter space has been explored by our models, there still remains further uncertainties such as mass transfer efficiency, pre-supernova core mass to remnant mass models or different wind prescriptions \citep{Broekgaarden:2021efa}. While exploring the entirety of 
this parameter space is beyond the scope of this study, more compact binary observations (having better signal-to-noise ratio that detects the masses and spins with better precision) will aid in restricting the uncertainties. 

We observed that all four models gave nearly identical conclusions, with Optimistic varying the most. Hence, while CE prescription does appear to affect the results, other variations within the limitations imposed by merger rates do not appear to dramatically alter our conclusions.

Future possible observations of more NS+BH mergers by LIGO/Virgo/KAGRA \citep{Belczynski:2016B, Mapelli:2018M, Broekgaarden:2021efa}, NS+BH binaries by LISA \citep{Chattopadhyay:2020lff, Wagg:2021} or even pulsars in binaries with BHs by radio telescopes like the SKA \citep{ Kyutoku2019K, Chattopadhyay:2020lff} will aid in shedding more light on the mass and spin distribution of these systems and hence put better constraints on massive binary evolution. Moreover, with more observations in gravitational waves as well as possible observations as radio pulsars of NS+BH systems, there will be higher scopes of differentiating the two NSBH and BHNS sub-populations. Table~\ref{tab:model_median} showing the medians of mass and spin from our different models and sub-populations can be used as a comparative grid for future observations of NS+BH coalescence(s).

\section*{Acknowledgements}

We thank Jarrod Hurley for the useful insights and discussions. DC thanks Michela Mapelli for her comments. We are grateful to the journal referee for their suggestions.
DC and SS are supported by the Australian Research Council (ARC) Centre of Excellence for Gravitational Wave Discovery (OzGrav), through project number CE170100004. DC is supported by the STFC grant ST/V005618/1. FA is supported by an STFC Rutherford fellowship (ST/P00492X/2).
SS is a recipient of an ARC Discovery Early Career Research Award (DE220100241).
This work made  use  of  the  OzSTAR  high  performance  computer at  Swinburne  University  of  Technology. 
OzSTAR  is funded by Swinburne University of Technology and the National Collaborative Research Infrastructure Strategy (NCRIS). 

\section*{Data Availability}

Simulations in this paper made use of the COMPAS rapid binary population synthesis code, which is freely available at \url{http://github.com/TeamCOMPAS/COMPAS} \citep{Stevenson:2017tfq,Vigna-Gomez:2018dza,COMPAS:2021methodsPaper}. 
The simulations performed in this work were simulated with a COMPAS version that predates the publicly available code. 
Our version of the code is most similar to version 02.13.01 of the publicly available COMPAS code. 
Requests for the original code can be made to the lead author. 




\bibliographystyle{mnras}
\bibliography{bib} 








\bsp	
\label{lastpage}
\end{document}